\documentclass{iopart}

\usepackage{graphicx}
\usepackage{float}
\usepackage{acronym}
\usepackage{color}
\usepackage{array}
\usepackage{psfrag}

\def\gsim{\mathrel{
 \rlap{\raise 0.511ex \hbox{$>$}}{\lower 0.511ex
 \hbox{$\sim$}}}}

\def\lsim{\mathrel{
 \rlap{\raise 0.511ex \hbox{$<$}}{\lower 0.511ex
 \hbox{$\sim$}}}}

\def\Msun{\ensuremath{M_{\odot}}}

\begin{document}

\acrodef{BBH}{binary black holes}
\acrodef{BNS}{binary neutron stars}
\acrodef{BHNS}{black hole-neutron star binaries}
\acrodef{SNR}{signal-to-noise ratio}
\acrodef{SPA}{stationary-phase approximation}
\acrodef{LIGO}{Laser Interferometer Gravitational-wave Observatory}
\acrodef{LSC}{LIGO Scientific Collaboration}
\acrodef{CBC}{compact binary coalescence}
\acrodef{GW}{gravitational waves}
\acrodef{ISCO}{innermost stable circular orbit}
\acrodef{FLSO}{frequency of last stable orbit}
\acrodef{PN}{post-Newtonian}
\acrodef{DFT}{discrete Fourier transform}
\acrodef{EOB}{effective one-body}
\acrodef{S4}{fourth science run}
\acrodef{S5}{fifth science run}
\acrodef{PSD}{power spectral density}
\acrodef{TT3}{Taylor T3}
\acrodef{FT}{Fourier transform}
\acrodef{TD}{time-domain}
\acrodef{FD}{frequency-domain}

\title[A tapering window for time-domain templates in 
gravitational wave searches]
{A tapering window for time-domain templates and simulated signals in
the detection of gravitational waves from 
coalescing compact binaries. \tiny{LIGO-P0900118-v4.}}
\author{DJA McKechan, C Robinson and BS Sathyaprakash}

\address{Queens Buildings, The Parade, Cardiff, CF24 3AA, UK}

\eads{\mailto{david.mckechan@astro.cf.ac.uk}, 
\mailto{craig.robinson@astro.cf.ac.uk}, 
\mailto{b.sathyaprakash@astro.cf.ac.uk}}

\begin{abstract}
Inspiral signals from binary black holes, in particular those with
masses in the range $10M_\odot \lsim M \lsim 1000 M_\odot,$ may last for 
only a few cycles within a detector's most sensitive frequency band. The 
spectrum of a square-windowed time-domain signal could contain unwanted 
power that can cause problems in gravitational wave data analysis, 
particularly when the waveforms are of
short duration. There may be leakage of power into 
frequency bins where no such power is expected, causing an excess of false
alarms. 
We present a method of tapering the time-domain waveforms that 
significantly reduces unwanted leakage of power, leading to a spectrum 
that agrees very well with that of a long duration signal. Our tapered 
window also decreases the false alarms caused by instrumental and 
environmental transients that are picked up by templates with spurious 
signal power. The suppression of background is an important goal in 
noise-dominated searches and can lead to an improvement in the detection 
efficiency of the search algorithms.
\end{abstract}

\submitto{\CQG}

\section{Introduction}\label{sec:intro}
Interferometric gravitational wave detectors are now operating
at sensitivity levels at which one can expect to detect inspirals
from compact binary coalescences at the rate of one per 5 years 
(optimistic rate) to one per 5000 years (pessimistic rate). When 
upgraded to advanced detectors, these might be as large as 400 per 
year to one per 2.5 yrs \cite{CBC:rates}. Even so, most of the 
inspiral signals are not likely to stand above the noise background.  
A variety of techiques to enhance signal visibility and reject
false alarms are currently being used in gravitational wave searches.
Examples include matched filtering for signals of known phase evolution 
\cite{BBCCS:2006}, wavelet transforms for transient signals of unknown 
shape \cite{Klimenko:2004a,Klimenko:2004b}, coherent search methods for burst 
signals \cite{ShourovLazzariniSteinSuttonSearleTinto2006}, etc. 
Moreover, vetoes based on the expected signal evolution \cite{Allen:2004} and
instrumental and environmental monitors \cite{Hanna:2006} have been developed 
over the past decade to improve detection probability and mitigate 
false alarms.  Detecting a signal buried in non-stationary noise 
is a challenging problem as some types of non-stationary noise
artefacts can partially mimic the signal.

Many of these techniques involve the computation of a correlation 
integral in which band-passed data are multiplied by the 
\ac{FD} model waveform or the \ac{DFT} 
of the \ac{TD} signal (see, for example, \cite{LIGOS1iul}).  
Here we consider a matched filtering search for inspiral signals 
where the \ac{DFT} of a \ac{TD} waveform is used to construct 
the correlation. A problem that has not been adequately addressed 
(see, however, Ref.\, \cite{Arnaud:2007a}) in this context is the 
effect of the window that is used in chopping a
\ac{TD} signal before computing its DFT.

Inevitably, all signal analysis algorithms use, implicitly or 
explicitly,  some form of window function.  An 
inspiral waveform sampled from a time when the signal's instantaneous 
frequency enters a detector's sensitive band until the time when it
reaches the \ac{FLSO} implicitly makes use of a square window.  
Signal analysis literature is full of examples of artefacts caused 
by the use of such window functions. For instance: leakage of power
from the main frequency bin where the signal is expected to lie into
neighbouring bins, loss of frequency resolution and corruption of
parameter estimation \cite{Windows:2009}.  In this paper
 we explore the problems caused by using a 
square window and suggest an alternative that cures some of the 
problems.

There is no unique, or favoured, windowing method. One is often
guided by the requirements of a particular analysis at hand.
In our case, a square window is especially bad since the leakage
of power outside the frequency range of interest can lead to
increased false alarm rate and poorer estimation of parameters.
One reason for increased false alarm rate could be that the
noise glitches in the detector look more like the untapered 
waveform and less like a tapered one.
We have explored the effect of a smoother window function, presented
in Section \ref{sec:method}, which has a far steeper fall-off of 
power outside the frequency range of interest.
Use of this window has cured several problems we had with a square
window. In Section \ref{sec:snr} we will discuss how tapering
helps in a more reliable signal spectral estimation and hence a
proper determination of the expected signal-to-noise ratio. Spectral
contamination is worse for larger mass black hole binaries as they
are in the detector's sensitive band for a shorter time and
the window function can only extend over a short time. 
It is for such signals that our tapered window offers the most improvement.
In Section \ref{sec:triggers} we will discuss how the rate of triggers
from a matched filtered search can vary depending on the kind of
window function used. We shall briefly mention in Section \ref{sec:pe}
what effect our window function has on parameter estimation, giving 
the conclusions of our study in Section~\ref{sec:conc}. 

\section{Window functions and their temporal and spectral characteristics}
\label{sec:method}

Let $h(t)$ denote a continuous differentiable function, for example
a gravitational wave signal emitted by a coalescing compact binary, and 
let $H(f)$ denote the \ac{FT} of $h(t)$ defined by
\begin{equation}
H(f) \equiv \int_{-\infty}^\infty h(t)\, \exp(2\pi ift)\, {\rm d}t.
\end{equation}
In reality the signal really does not last for an infinite time. The
\ac{FT} of a signal of finite duration lasting, say, 
from $-{T}/{2}$ to ${T}/{2},$ can be represented either by using the 
limits of the integral to go from $-{T}/{2}$ to ${T}/{2}$ or by using 
a window function. The latter is preferred so as to preserve the 
definition of the \ac{FT}. 

A window function is a function that has either a finite 
support or falls off sufficiently rapidly as $t\rightarrow \pm \infty.$ 
Two simple windows that have finite support are the square window 
$s_T(t)$ defined by 
\begin{eqnarray}
s_T(t) = \left \{ \begin{array}{ccc} 
1, & \quad & -\frac{T}{2} \le t \le \frac{T}{2}\\
0, & \quad & \rm otherwise,\\
\end{array}
\right .
\end{eqnarray}
and the triangular window $b_T(t)$ defined by 
\begin{eqnarray}
b_T(t) = \left \{ \begin{array}{ccc} 
(1-2|t|/T), & \quad & -\frac{T}{2} \le t \le \frac{T}{2}\\
0, & \quad & \rm otherwise.\\
\end{array}
\right .
\end{eqnarray}
Neither the square nor the triangular window are 
differentiable everywhere. 
As a result, they are not functions of finite bandwidth. 
In other words, their \ac{FT}s, $S_T(f)$ and $B_T(f)$, do not have finite 
support in the \ac{FD}: $|S(f)| > 0$ for $-\infty \le f \le \infty.$ 
In the case of a square window the \ac{FT} $S(f)$ is a 
sinc-function, $|S_T(f)| = T {\rm sinc}(\pi f T),$ which is
peaked at $f=0,$ with a width $\pi/T$ and falls off as $f^{-1}$ as 
$f \rightarrow \pm \infty.$ The lack of finite support in the Fourier
domain could sometimes cause problems, especially when the width of 
the window in \ac{TD} is too small. For functions
that have infinite bandwidth the sampling theorem does not hold but
this is not a serious drawback if the \ac{FT} falls off sufficiently
fast above the Nyquist frequency. However, there could be other
issues when the window leads to leakage of power outside a region
of interest as we shall see below.

\subsection{The Planck-taper window function}
A signal $h(t)$ with the window $w_T(t)$ applied to it, in other words
the windowed signal $h_w(t),$ is defined by
\begin{equation}
\label{eq:hw}
h_w(t) \equiv h(t) w_T(t).
\end{equation}
The convolution theorem states that the \ac{FT} of the product of
two functions $h(t)$ and $w_T(t)$ is the convolution of individual
\ac{FT}s:
\begin{eqnarray}
\label{eq:conv}
H_w(f) & = \int_{-\infty}^\infty h(t) w_T(t) \exp(2\pi i f t) {\rm d}t \\
\label{eq:conv2}
 & = H(f) * W_T(f) = \int_{-\infty}^\infty H(f') W_T(f-f') {\rm d}f'.
\end{eqnarray}
We can now see why a window whose power in the \ac{FD} does not 
fall off sufficiently rapidily
might be problematic. The convolution integral will have contributions
from all frequencies. Suppose we are interested in matched filtering the
data with an inspiral signal from a compact coalescing binary whose 
instantaneous frequency varies from $f_a$ at time $t_a$ to $f_b$ at
time $t_b$. One would normally achieve this by using a square window 
$s_T(t)$ that is centered at $(t_a+t_b)/2$ with width $T=t_b-t_a.$ However,
we can see from (\ref{eq:conv2}) that the convolution integral will
have contributions from outside the frequency range of interest.

\begin{figure*}
\begin{minipage}[h]{4cm}
%
%
\begin{psfrags}%
\psfragscanon%
%
\psfrag{s01}[t][t]{\color[rgb]{0,0,0}\setlength{\tabcolsep}{0pt}\begin{tabular}{c}Time /s\end{tabular}}%
\psfrag{s06}[][]{\color[rgb]{0,0,0}\setlength{\tabcolsep}{0pt}\begin{tabular}{c} \end{tabular}}%
\psfrag{s07}[][]{\color[rgb]{0,0,0}\setlength{\tabcolsep}{0pt}\begin{tabular}{c} \end{tabular}}%
\psfrag{s28}[t][t]{\color[rgb]{0,0,0}\setlength{\tabcolsep}{0pt}\begin{tabular}{c}$w_T(t;1/10)$\end{tabular}}%
\psfrag{s29}[][]{\color[rgb]{0,0,0}\setlength{\tabcolsep}{0pt}\begin{tabular}{c}$w_T(t;1/30)\ \ \ $\end{tabular}}%
\psfrag{s30}[][]{\color[rgb]{0,0,0}\setlength{\tabcolsep}{0pt}\begin{tabular}{c}$w_T(t;1/100)\ \ \ $\end{tabular}}%
\psfrag{s31}[][]{\color[rgb]{0,0,0}\setlength{\tabcolsep}{0pt}\begin{tabular}{c}$s_T(t)\ \ \ $\end{tabular}}%
%
\psfrag{x01}[t][t]{0}%
\psfrag{x02}[t][t]{0.1}%
\psfrag{x03}[t][t]{0.2}%
\psfrag{x04}[t][t]{0.3}%
\psfrag{x05}[t][t]{0.4}%
\psfrag{x06}[t][t]{0.5}%
\psfrag{x07}[t][t]{0.6}%
\psfrag{x08}[t][t]{0.7}%
\psfrag{x09}[t][t]{0.8}%
\psfrag{x10}[t][t]{0.9}%
\psfrag{x11}[t][t]{1}%
\psfrag{x12}[t][t]{0.8}%
\psfrag{x13}[t][t]{0.85}%
\psfrag{x14}[t][t]{0.9}%
\psfrag{x15}[t][t]{0.95}%
\psfrag{x16}[t][t]{1}%
%
\psfrag{v01}[r][r]{0}%
\psfrag{v02}[r][r]{0.1}%
\psfrag{v03}[r][r]{0.2}%
\psfrag{v04}[r][r]{0.3}%
\psfrag{v05}[r][r]{0.4}%
\psfrag{v06}[r][r]{0.5}%
\psfrag{v07}[r][r]{0.6}%
\psfrag{v08}[r][r]{0.7}%
\psfrag{v09}[r][r]{0.8}%
\psfrag{v10}[r][r]{0.9}%
\psfrag{v11}[r][r]{1}%
\psfrag{v12}[r][r]{0}%
\psfrag{v13}[r][r]{0.2}%
\psfrag{v14}[r][r]{0.4}%
\psfrag{v15}[r][r]{0.6}%
\psfrag{v16}[r][r]{0.8}%
\psfrag{v17}[r][r]{1}%
\psfrag{v18}[r][r]{1.2}%
%
\resizebox{6cm}{!}{\includegraphics{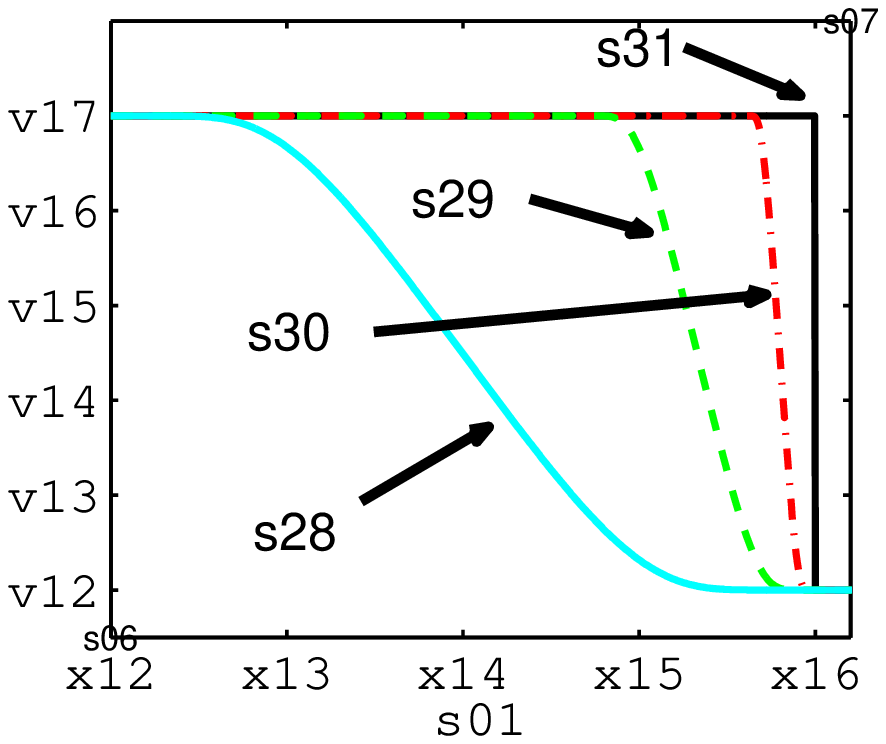}}%
\end{psfrags}%
%

\end{minipage}
\hspace{2.25cm}
\begin{minipage}[h]{4cm}
%
%
\begin{psfrags}%
\psfragscanon%
%
\psfrag{s02}[t][t]{\color[rgb]{0,0,0}\setlength{\tabcolsep}{0pt}\begin{tabular}{c}Frequency /Hz\end{tabular}}%
\psfrag{s17}[t][t]{\color[rgb]{0,0,0}\setlength{\tabcolsep}{0pt}\begin{tabular}{c}$W_T(f;1/10)$\end{tabular}}%
\psfrag{s18}[b][b]{\color[rgb]{0,0,0}\setlength{\tabcolsep}{0pt}\begin{tabular}{c}$W_T(f;1/30)$\end{tabular}}%
\psfrag{s19}[b][b]{\color[rgb]{0,0,0}\setlength{\tabcolsep}{0pt}\begin{tabular}{c}$S_T(f)$\end{tabular}}%
\psfrag{s20}[b][b]{\color[rgb]{0,0,0}\setlength{\tabcolsep}{0pt}\begin{tabular}{c}$W_T(f;1/100)$\end{tabular}}%
%
\psfrag{x01}[t][t]{0}%
\psfrag{x02}[t][t]{0.1}%
\psfrag{x03}[t][t]{0.2}%
\psfrag{x04}[t][t]{0.3}%
\psfrag{x05}[t][t]{0.4}%
\psfrag{x06}[t][t]{0.5}%
\psfrag{x07}[t][t]{0.6}%
\psfrag{x08}[t][t]{0.7}%
\psfrag{x09}[t][t]{0.8}%
\psfrag{x10}[t][t]{0.9}%
\psfrag{x11}[t][t]{1}%
\psfrag{x12}[b][b]{\color[rgb]{0,0,0}\setlength{\tabcolsep}{0pt}\begin{tabular}{c}$10^0$\end{tabular}}%
\psfrag{x13}[b][b]{\color[rgb]{0,0,0}\setlength{\tabcolsep}{0pt}\begin{tabular}{c}$10^1$\end{tabular}}%
\psfrag{x14}[b][b]{\color[rgb]{0,0,0}\setlength{\tabcolsep}{0pt}\begin{tabular}{c}$10^2$\end{tabular}}%
\psfrag{x15}[b][b]{\color[rgb]{0,0,0}\setlength{\tabcolsep}{0pt}\begin{tabular}{c}$10^3$\end{tabular}}%
%
\psfrag{v01}[r][r]{0}%
\psfrag{v02}[r][r]{0.1}%
\psfrag{v03}[r][r]{0.2}%
\psfrag{v04}[r][r]{0.3}%
\psfrag{v05}[r][r]{0.4}%
\psfrag{v06}[r][r]{0.5}%
\psfrag{v07}[r][r]{0.6}%
\psfrag{v08}[r][r]{0.7}%
\psfrag{v09}[r][r]{0.8}%
\psfrag{v10}[r][r]{0.9}%
\psfrag{v11}[r][r]{1}%
\psfrag{v12}[b][b]{\color[rgb]{0,0,0}\setlength{\tabcolsep}{0pt}\begin{tabular}{c}$10^{-10}$\end{tabular}}%
\psfrag{v13}[b][b]{\color[rgb]{0,0,0}\setlength{\tabcolsep}{0pt}\begin{tabular}{c}$10^{-8}$\end{tabular}}%
\psfrag{v14}[b][b]{\color[rgb]{0,0,0}\setlength{\tabcolsep}{0pt}\begin{tabular}{c}$10^{-6}$\end{tabular}}%
\psfrag{v15}[b][b]{\color[rgb]{0,0,0}\setlength{\tabcolsep}{0pt}\begin{tabular}{c}$10^{-4}$\end{tabular}}%
\psfrag{v16}[b][b]{\color[rgb]{0,0,0}\setlength{\tabcolsep}{0pt}\begin{tabular}{c}$10^{-2}$\end{tabular}}%
\psfrag{v17}[b][b]{\color[rgb]{0,0,0}\setlength{\tabcolsep}{0pt}\begin{tabular}{c}$$\end{tabular}}%
%
\resizebox{6cm}{!}{\includegraphics{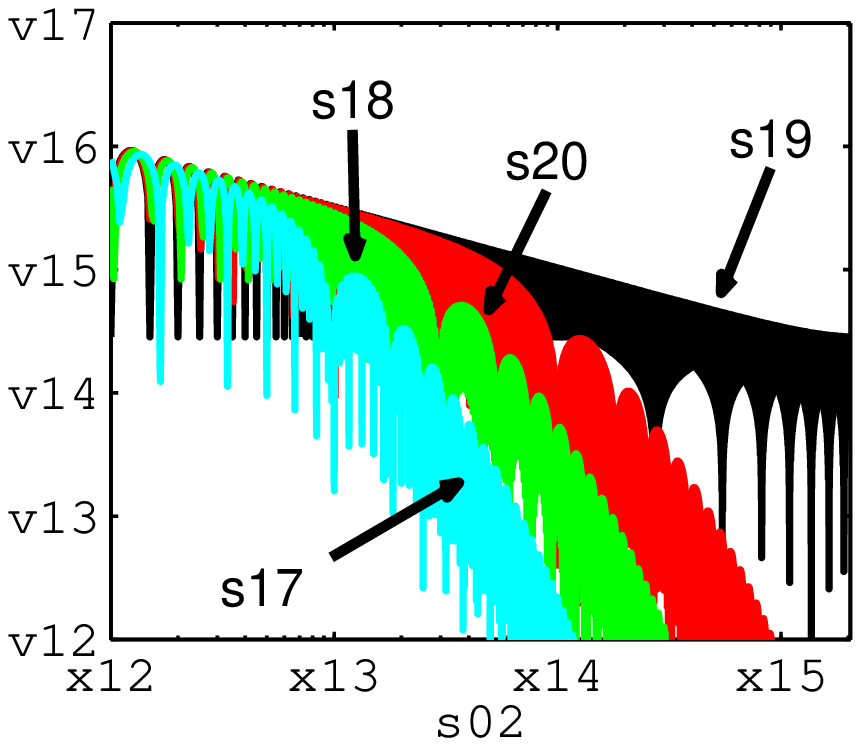}}%
\end{psfrags}%
%

\end{minipage}
\caption{The Planck-taper window in the \ac{TD} (left), for three 
different choices of the parameter $\epsilon=0.01,\,0.033,\,0.1,$ and their
power spectra (right). For reference we have included the
square window with the same effective width as the Planck-taper window. 
}
\label{fig:window}
\end{figure*}

To circumvent this problem we propose to use a window function 
that falls off rapidly outside the frequency range of interest.
Inspired by the tapering function used in Damour et al 
\cite{Damour:2000gg} we define a new window function $\sigma(t)$ by
\begin{eqnarray}
\label{eq:taper}
\fl \sigma_T(t; \epsilon) = 
\left \{ \begin{array}{lllll}
 0, & \quad & & t \le t_1 & t_1 = -\frac{T}{2}, \\[0.1cm]
 \frac{1}{\exp(z(t))+1}, & \quad & z(t)=\frac{t_2-t_1}{t-t_1}+
 \frac{t_2-t_1}{t-t_2}, & t_1  < t < t_2,  & t_2 = -\frac{T}{2}(1-2\epsilon)\\[0.1cm]
 1, & \quad & \quad & t_2 \le t \le t_3, & t_3 = \frac{T}{2}(1-2\epsilon) \\[0.1cm]
 \frac{1}{\exp(z(t))+1}, & \quad & z(t)=\frac{t_3-t_4}{t-t_3}+
 \frac{t_3-t_4}{t-t_4}, & t_3  < t < t_4, & t_4=\frac{T}{2}, \\[0.1cm]
 0, & \quad & & t_4 \le t.
 \end{array}
 \right .
\end{eqnarray}
Here $T$ is the width of the window and $\epsilon$ is the fraction of the 
window width over which the window function smoothly rises from 0 at $t=t_1$ 
to 1 at $t=t_2$ or falls from 1 at $t=t_3$ to 0 at $t=t_4.$
We shall call $\sigma(t)$ the \textit{Planck-taper window} as 
the basic functional
form is that of the Planck distribution. The motivation for choosing this 
window function is to reduce the leakage of power in the \ac{FD} 
but at the same time not to lose too much of the length of the signal in the 
\ac{TD}. The choice of $\epsilon$ will affect both aspects significantly.
In figure \ref{fig:window} we have shown the window function for several 
choices of the parameter $\epsilon=0.01, 0.033, 0.1.$.
At lower frequencies the spectrum of the Planck-taper window falls 
off at the same rate (i.e., $1/f$) as a square window. But beyond a
certain frequency $f_0 \sim (\epsilon T)^{-1},$ the spectrum falls off 
far faster.

A key feature in our use of the Planck-taper window is the
automated and waveform-dependent adjustment of $\epsilon$ as 
discussed in section~\ref{sec:imp} below.

\subsection{Implementation of the window}
\label{sec:imp}
\begin{figure}[ht]
\begin{minipage}[h]{4cm}
%
%
\begin{psfrags}%
\psfragscanon%
%
\psfrag{s03}[t][t]{\color[rgb]{0,0,0}\setlength{\tabcolsep}{0pt}\begin{tabular}{c}Array index\end{tabular}}%
%
\psfrag{x01}[t][t]{0}%
\psfrag{x02}[t][t]{20}%
\psfrag{x03}[t][t]{40}%
\psfrag{x04}[t][t]{60}%
\psfrag{x05}[t][t]{80}%
\psfrag{x06}[t][t]{100}%
\psfrag{x07}[t][t]{120}%
\psfrag{x08}[t][t]{140}%
\psfrag{x09}[t][t]{160}%
\psfrag{x10}[t][t]{180}%
%
\psfrag{v01}[r][r]{-3}%
\psfrag{v02}[r][r]{-2}%
\psfrag{v03}[r][r]{-1}%
\psfrag{v04}[r][r]{0}%
\psfrag{v05}[r][r]{1}%
\psfrag{v06}[r][r]{2}%
\psfrag{v07}[r][r]{3}%
%
\resizebox{6.5cm}{!}{\includegraphics{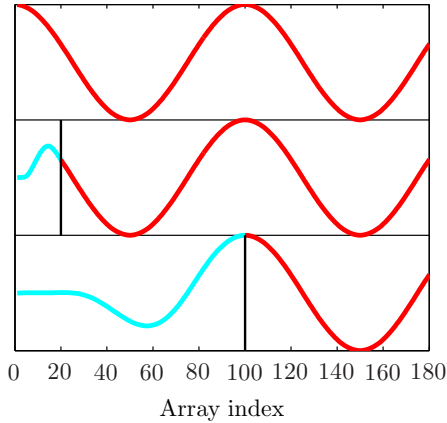}}%
\end{psfrags}%
%

\end{minipage}
\hspace{1.25cm}
\begin{minipage}[h]{7.5cm}
\caption{
The window function has been applied to the start of a cosine wave (top curve) 
using two methods. In the first case it is applied from $j=1$ up to an
arbitrary choice of $j=20$ (middle), whereas in the second case it is applied 
up to the second maximum at $j=100$ (bottom). 
The lighter coloured parts of the middle and bottom curves (to the left of the
black vertical lines) show where the taper has been applied.}
\label{fig:taperapplication}
\end{minipage}
\end{figure}

We discretise (\ref{eq:taper}) by replacing $t,t_1,t_2, t_3, t_4$ 
with the array indices $j,j_1,j_2, j_3, j_4.$ 
In this notation the parameter epsilon is approximated by
$\epsilon \simeq (j_2-j_1)/N,$ 
where $N$ is the number of data points in the waveform.  
The start and end of the 
waveform are denoted by $j_1$ and $j_4$, respectively. 
The values of $j_2$
and $j_3$ have to be chosen judiciously to avoid leakage of power.
We choose $j_2$ and $j_3$ to be the array index corresponding
to the second stationary point after $j_1$ and before $j_4$ 
(see figure~\ref{fig:taperapplication}). Applying the transition stage 
of $\sigma$ from a crest/trough ensures that the window does not 
have a sudden impact
on the behaviour of the waveform. The first stationary point would not be an 
appropriate choice as it may occur within only a few array points of
$j_1$ or $j_4,$ causing $\epsilon$ to be too small. One could 
choose the 3rd, 4th or 5th, but using such later maxima would reduce 
the genuine power of the waveform more than what might be acceptable.

\subsection{Comparison with other windows}
We do not compare the performance of Planck-taper with other
commonly used windows, e.g., Bartlet, Hann or Welch. Such windows 
transition between $0$ and $1$ 
over $j=1,\ldots,N/2$, where the window is of length $N$, 
producing significant differences between $h(t)$ and $h_w(t)$ in (\ref{eq:hw}).
The power is therefore supressed at the beginning and end of $h(t)$.
This is acceptable when computing the \ac{PSD} of a data segment,
but would cause a problem if applied to a template waveform as the phase (frequency)
and amplitude of $h(t)$ are both instantaneous functions of $t$,
with the most power at the end of the waveform. 

Windows with properties similar to Planck-taper, such as having a central flat 
region, do exist. For example, the Tukey window~\cite{1978IEEE}, which has been 
used in gravitational-wave data analysis recently~\cite{2007PhRvD}, may offer
a good comparison. However, a key feature in our study of the Planck-taper
window is the waveform-dependent adjustment of $j_2$ and $j_3$. Whilst this
automation could be considered separately from the Planck-taper window and
used on other windows defined by the points $j_{1,2,3,4}$, we have not
done so here. Given the shared features of the Tukey window with 
Planck-Taper one might expect similar results.

\section{Effect of the window function on the signal spectrum}
In this section we will examine the power spectrum of
the waveform of a coalescing binary emitted during the inspiral phase.
The waveforms are modelled using the \ac{PN} approximation. However,
even within the \ac{PN} approximation, there are several different
ways in which one might construct the waveform 
\cite{Damour:2000zb,Buonanno:2009zt}. 
Two such models widely used in the search for compact binary
coalescences are \ac{TT3} and the \ac{SPA}.  \ac{TT3} is a \ac{TD} 
signal model in which the amplitude and phase of the signal are both
explicit functions of time. In the so-called restricted \ac{PN} approximation
the signal consists of the dominant harmonic at twice the orbital
frequency, but not higher order \ac{PN} corrections consisting of other
harmonics, and the phase is a \ac{PN} expansion that is currently known
to ${\cal O}(v^7)$ in the expansion parameter $v$ -- the relative velocity
of the two stars. The \ac{SPA} is the Fourier transform of the \ac{TT3} 
model obtained
by using the stationary phase approximation to the Fourier integral
\cite{SathyaDhurandhar:1991}.  A template belonging to the \ac{TT3} model
is defined for times when the gravitational wave frequency is within the 
detector's sensitivity band until it reaches \ac{FLSO}. This means one is
in effect multiplying a square window with a continuous function. 

\begin{figure}[ht]
\hspace{0.35cm}
\begin{minipage}[h]{6cm}
%
%
\begin{psfrags}%
\psfragscanon%
%
\psfrag{s05}[t][t]{\color[rgb]{0,0,0}\setlength{\tabcolsep}{0pt}\begin{tabular}{c}Frequency /Hz\end{tabular}}%
\psfrag{s06}[b][b]{\color[rgb]{0,0,0}\setlength{\tabcolsep}{0pt}\begin{tabular}{c}$f.|H(f)|^2 / S_h(f)$\end{tabular}}%
\psfrag{s07}[b][b]{\color[rgb]{0,0,0}\setlength{\tabcolsep}{0pt}\begin{tabular}{c}$20\ M_\odot$\end{tabular}}%
\psfrag{s10}[][]{\color[rgb]{0,0,0}\setlength{\tabcolsep}{0pt}\begin{tabular}{c} \end{tabular}}%
\psfrag{s11}[][]{\color[rgb]{0,0,0}\setlength{\tabcolsep}{0pt}\begin{tabular}{c} \end{tabular}}%
\psfrag{s12}[l][l]{\color[rgb]{0,0,0}SPA}%
\psfrag{s13}[l][l]{\color[rgb]{0,0,0}H$_S$}%
\psfrag{s14}[l][l]{\color[rgb]{0,0,0}H$_\sigma$}%
\psfrag{s15}[l][l]{\color[rgb]{0,0,0}SPA}%
%
\psfrag{x01}[t][t]{$10^2$}%
\psfrag{x02}[t][t]{$10^3$}%
%
\psfrag{v01}[r][r]{$10^{-8}$}%
\psfrag{v02}[r][r]{$10^{-7}$}%
\psfrag{v03}[r][r]{$10^{-6}$}%
\psfrag{v04}[r][r]{$10^{-5}$}%
\psfrag{v05}[r][r]{$10^{-4}$}%
\psfrag{v06}[r][r]{$10^{-3}$}%
%
\resizebox{6.5cm}{!}{\includegraphics{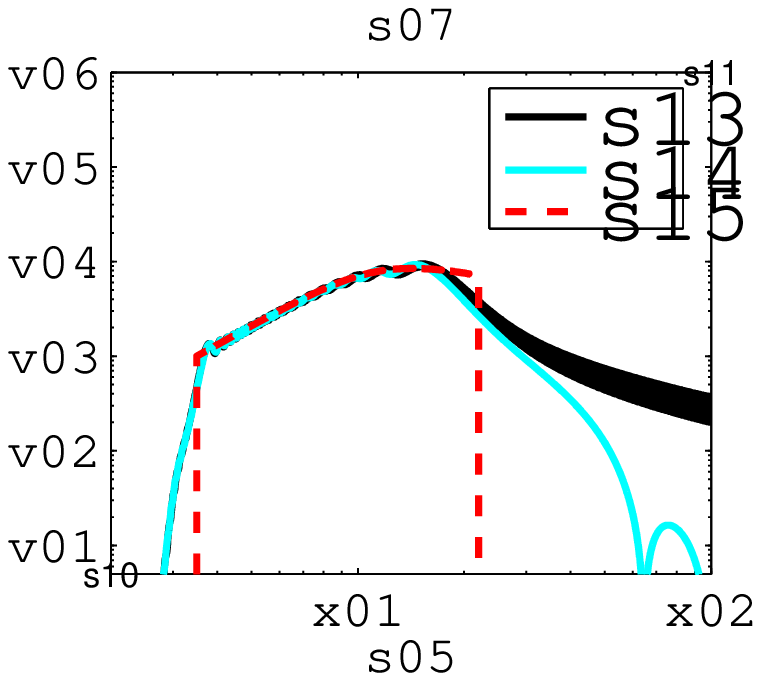}}%
\end{psfrags}%
%

\end{minipage}
\hspace{0.1cm}
\begin{minipage}[h]{6cm}
%
%
\begin{psfrags}%
\psfragscanon%
%
\psfrag{s05}[t][t]{\color[rgb]{0,0,0}\setlength{\tabcolsep}{0pt}\begin{tabular}{c}Frequency /Hz\end{tabular}}%
\psfrag{s06}[b][b]{\color[rgb]{0,0,0}\setlength{\tabcolsep}{0pt}\begin{tabular}{c}$$\end{tabular}}%
\psfrag{s07}[b][b]{\color[rgb]{0,0,0}\setlength{\tabcolsep}{0pt}\begin{tabular}{c}$80\ M_\odot$\end{tabular}}%
\psfrag{s10}[][]{\color[rgb]{0,0,0}\setlength{\tabcolsep}{0pt}\begin{tabular}{c} \end{tabular}}%
\psfrag{s11}[][]{\color[rgb]{0,0,0}\setlength{\tabcolsep}{0pt}\begin{tabular}{c} \end{tabular}}%
\psfrag{s12}[l][l]{\color[rgb]{0,0,0}SPA}%
\psfrag{s13}[l][l]{\color[rgb]{0,0,0}H$_S$}%
\psfrag{s14}[l][l]{\color[rgb]{0,0,0}H$_\sigma$}%
\psfrag{s15}[l][l]{\color[rgb]{0,0,0}SPA}%
%
\psfrag{x01}[t][t]{$10^2$}%
\psfrag{x02}[t][t]{$10^3$}%
%
\psfrag{v01}[r][r]{$10^{-8}$}%
\psfrag{v02}[r][r]{$10^{-7}$}%
\psfrag{v03}[r][r]{$10^{-6}$}%
\psfrag{v04}[r][r]{$10^{-5}$}%
\psfrag{v05}[r][r]{$10^{-4}$}%
\psfrag{v06}[r][r]{$10^{-3}$}%
%
\resizebox{6.5cm}{!}{\includegraphics{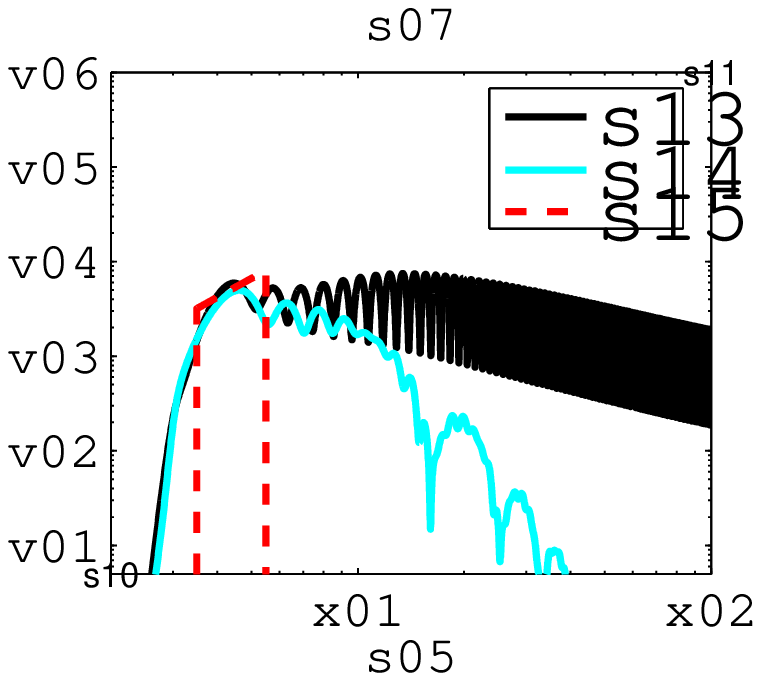}}%
\end{psfrags}%
%

\end{minipage}
\caption{The plots show the \ac{SNR} integrand, where the waveform
is generated from a frequency of $35$ Hz to the \ac{FLSO} of the source, 
computed using the initial
LIGO sensitivity for sources of total mass  $20\,\Msun$  
and $80\Msun.$ In the case of the Planck-taper window the \ac{SNR}
integrand falls off far faster than in the case of the square window
or the \ac{TT3} model.}
\label{fig:spectra}
\end{figure}

Figure~\ref{fig:spectra} shows the \ac{SNR} integrand of the
\ac{SPA}, computed using the initial \ac{LIGO} design 
\ac{PSD}~\cite{Damour:2000zb}. The inspiral waveform 
is defined from a lower cutoff frequency of $35\,\rm Hz$ up to its \ac{FLSO},
for $20\, \Msun$ and $80\,\Msun$ equal-mass binaries. 
The \ac{DFT} of the \ac{TT3}, generated between the same frequencies, 
with a square window (or rather no window), labelled H$_S$, and 
with the Planck-taper window, labelled H$_\sigma$, are also plotted.
Where the Planck-taper window is used the excess power, 
that above \ac{FLSO}, decreases rapidly and the spectrum is closer
to that of the \ac{SPA}.

\section{Effect of the window function on the estimation of the 
signal-to-noise ratio}
\label{sec:snr}

Gravitational wave searches for known signals, such as those emitted from 
a \ac{CBC} \cite{Collaboration:2009tt,Abbott:2009qj},
rely upon signal models for two primary reasons. Firstly, they are used as 
templates to matched filter the data. Secondly, they are injected into the data 
as simulated signals to estimate the efficiency of the detector to detect
such signals.  If the signal/template models are 
generated in the \ac{TD} then they must undergo a \ac{DFT} if the data
are analysed in the \ac{FD} as is the case for the current \ac{LIGO}
matched filter code.

The expectation value for the \ac{SNR} of a signal in stationary Gaussian 
noise, when the signal and template match exactly, is given by 
\begin{equation}
\label{eq:snr}
\left<\rho^2\right> = 4 
\int_{f_{lower}}^{f_{nyquist}}df\frac{{|H(f)|}^2}{S_h(f)}
\simeq 4\Delta f \sum_{k=1}^{N/2-1}\frac{{|H_k|}^2}{S_{hk}},
\end{equation}
where $S_h(f)$ is the detector noise \ac{PSD} and $f_{lower}$ is 
the detector lower cutoff frequency chosen so that the contribution
to the \ac{SNR} integral from frequencies $f<f_{lower}$ is negligible.
The second of the expressions 
on the RHS is a discretised evaluation of the \ac{SNR} which is 
often used in numerical calculations. Here $H_k,$ $k=0,\ldots,N/2,$  
is the \ac{DFT} of the signal defined for positive frequencies
and $S_{hk}$ is the discretised \ac{PSD}.

The amplitude of an inspiral signal increases with the total mass of 
the system; conversely, the \ac{FLSO} of the signal is inversely 
proportional to the total mass.  Therefore,
as the total mass of a system increases, the amplitude of the signal and
the \ac{FLSO} will have opposing effects.
For lower mass systems, the increasing amplitude causes the \ac{SNR}
to increase as a function of the total mass. However, for higher mass
systems, the reduction in the \ac{FLSO} causes the signal to have less
power in band. As a result, the \ac{SNR} will decrease as a function 
of the total mass. The relatively low \ac{FLSO} of the higher mass templates, 
coupled with their short duration, lead them to be particularly 
susceptible to artefacts of spectral leakage in the DFT.

\begin{figure}[ht]
\hspace{0.5cm}
\begin{minipage}[h]{4cm}
%
%
\begin{psfrags}%
\psfragscanon%
%
\psfrag{s05}[t][t]{\color[rgb]{0,0,0}\setlength{\tabcolsep}{0pt}\begin{tabular}{c}Total Mass / $M_\odot$\end{tabular}}%
\psfrag{s06}[b][b]{\color[rgb]{0,0,0}\setlength{\tabcolsep}{0pt}\begin{tabular}{c}SNR\end{tabular}}%
\psfrag{s10}[][]{\color[rgb]{0,0,0}\setlength{\tabcolsep}{0pt}\begin{tabular}{c} \end{tabular}}%
\psfrag{s11}[][]{\color[rgb]{0,0,0}\setlength{\tabcolsep}{0pt}\begin{tabular}{c} \end{tabular}}%
\psfrag{s12}[l][l]{\color[rgb]{0,0,0}Tapered}%
\psfrag{s13}[l][l]{\color[rgb]{0,0,0}Untapered}%
\psfrag{s14}[l][l]{\color[rgb]{0,0,0}Tapered}%
%
\psfrag{x01}[t][t]{0}%
\psfrag{x02}[t][t]{0.1}%
\psfrag{x03}[t][t]{0.2}%
\psfrag{x04}[t][t]{0.3}%
\psfrag{x05}[t][t]{0.4}%
\psfrag{x06}[t][t]{0.5}%
\psfrag{x07}[t][t]{0.6}%
\psfrag{x08}[t][t]{0.7}%
\psfrag{x09}[t][t]{0.8}%
\psfrag{x10}[t][t]{0.9}%
\psfrag{x11}[t][t]{1}%
\psfrag{x12}[t][t]{0}%
\psfrag{x13}[t][t]{20}%
\psfrag{x14}[t][t]{40}%
\psfrag{x15}[t][t]{60}%
\psfrag{x16}[t][t]{80}%
\psfrag{x17}[t][t]{100}%
\psfrag{x18}[t][t]{120}%
%
\psfrag{v01}[r][r]{0}%
\psfrag{v02}[r][r]{0.1}%
\psfrag{v03}[r][r]{0.2}%
\psfrag{v04}[r][r]{0.3}%
\psfrag{v05}[r][r]{0.4}%
\psfrag{v06}[r][r]{0.5}%
\psfrag{v07}[r][r]{0.6}%
\psfrag{v08}[r][r]{0.7}%
\psfrag{v09}[r][r]{0.8}%
\psfrag{v10}[r][r]{0.9}%
\psfrag{v11}[r][r]{1}%
\psfrag{v12}[r][r]{4}%
\psfrag{v13}[r][r]{6}%
\psfrag{v14}[r][r]{8}%
\psfrag{v15}[r][r]{10}%
\psfrag{v16}[r][r]{12}%
\psfrag{v17}[r][r]{14}%
\psfrag{v18}[r][r]{16}%
\psfrag{v19}[r][r]{18}%
%
\resizebox{7cm}{!}{\includegraphics{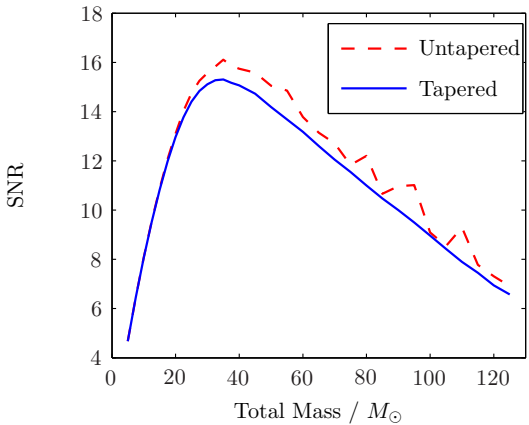}}%
\end{psfrags}%
%

\end{minipage}
\hspace{0.75cm}
\begin{minipage}[h]{7.5cm}
\caption{
The \ac{SNR} vs the total mass of the source for signals corresponding to 
compact binary systems directly overhead a detector of initial LIGO design 
\ac{PSD}.  We plot the \ac{SNR} obtained using the \ac{DFT} of \ac{TD} 
waveforms with a square window (dashed curve) and with the Planck-taper window 
(solid curve). Here the systems are overhead the detector at an effective
distance of $65Mpc$, using a fixed mass ratio of $5:1$ and a fixed
inclination angle of $45^o$.}
\label{fig:snrvsmass}
\end{minipage}
\end{figure}

Figure~\ref{fig:snrvsmass} shows the \ac{SNR} for \ac{TT3} inspiral 
waveforms that are $2$\ac{PN} in amplitude and phase, plotted as a 
function of the total mass for two choices
of the window function: the dashed curve corresponds to the
square window and the solid curve to the Planck-taper window. All
other parameters are the same in both cases.  When the Planck-taper window
is used, the curve exhibits the expected behaviour, whereas in the
case of a square window , the \ac{SNR} curve is `jagged' which is unexpected
given that stationary Gaussian noise was used in the estimation of the
\ac{SNR}. This behaviour is most likely explained by the excess power from the 
\ac{DFT} of the waveform. 

It should be noted that integrating to \ac{FLSO} rather than Nyquist in Eq.\
(\ref{eq:snr}), is not considered appropriate here. Firstly, the higher 
harmonics in the amplitude corrected waveforms contain power above \ac{FLSO}, 
(which becomes more signifcant for high mass systems).
Secondly, cutting off the integration at \ac{FLSO} is essentially 
the application of a square window to the template waveform in the
frequency domain. This will lead to leakage of power in the time domain
which is not a desirable feature. The problem
of using a square-windowed \ac{TD} template as our matched filter is not
that there is power above \ac{FLSO}; it is that the excess power in this
region, present due to windowing, but not present in a genuine signal will 
lead to unnecessary false alarms in a search.

\section{Effect of window functions on Trigger Rates}\label{sec:triggers}
To assess the effect that tapering of templates has on trigger rates, we have
applied the \ac{LSC} \ac{CBC} pipeline~\cite{LIGOS3S4Tuning,LIGOS3S4all,
Collaboration:2009tt,Abbott:2009qj,Allen:2005fk} 
to data taken during the \ac{S4} of the \ac{LIGO},
which took place from February 22 - March 23, 2005. The basic topology of
the pipeline is similar to that used in many previous 
searches~\cite{LIGOS3S4all,Collaboration:2009tt,Abbott:2009qj}, and consists of the following main steps:
\begin{itemize}
\item The template bank is chosen such that the loss of SNR due to having a 
finite number of templates is no more than $3\%$ for any signal belonging
to a given family of waveforms~\cite{hexabank,BBCCS:2006}.
\item Matched filter the data with the generated templates. A trigger is
generated at times when the SNR is larger than a given threshold. The output
of this stage is a list of \textit{first-stage single-detector triggers}.
\item Check for coincident events between different detectors. For an event
to be deemed coincident, the parameters seen in at least two detectors (for
instance, the masses of the system, the time of coalescence, \ldots) 
should agree to within a certain tolerance \cite{Robinson:2008}. 
The output of this stage is a list of \textit{first-stage coincident triggers}.
\item Re-filter the data using only templates associated with coincident 
triggers. This time, the triggers are subjected to further 
signal-based vetoes, some of which are computationally costly, such as the 
chi-squared veto~\cite{Allen:2004}. This produces a list of 
\textit{second-stage single-detector triggers}.
\item Check for coincident events between detectors using the second-stage 
single-detector triggers. This produces a list of \textit{second-stage coincident triggers}.
\end{itemize}

In this study the data were filtered using the \ac{EOB} templates 
\cite{BuonannoDamour:1999,BuonannoDamour:2000,Damour:2000zb}, tuned to recent 
results in numerical relativity~\cite{Buonanno:2007pf,Damour:2007yf}, 
with a total mass in the range $25-100M_\odot$. This choice agrees with
the templates used to search for signals from high-mass \ac{CBC}s in 
data from LIGO's \ac{S5}. Because the \ac{EOB} waveforms used as templates 
contain the inspiral, merger and ringdown phases, 
there was no need to taper the end of the waveform. 
Therefore, in this case, the taper specified in (\ref{eq:taper}) was only 
applied to the start of the waveform. Although this may reduce the effect 
the taper has in comparison to tapering both ends of an inspiral-only template, 
it is of more interest to evaluate the performance in a realistic search case.
It should be noted that the tapering window is 
\textit{explicitly applied to the
template waveform} where the length of the waveform is less than the
length of the data segment that is matched filtered. We do not apply
any window to the data segment.

\begin{figure}[ht]
\begin{minipage}[h]{4cm}
%
%
\begin{psfrags}%
\psfragscanon%
%
\psfrag{s04}[][]{\color[rgb]{0,0,0}\setlength{\tabcolsep}{0pt}\begin{tabular}{c} \end{tabular}}%
\psfrag{s05}[][]{\color[rgb]{0,0,0}\setlength{\tabcolsep}{0pt}\begin{tabular}{c} \end{tabular}}%
\psfrag{s06}[b][b]{\color[rgb]{0,0,0}\setlength{\tabcolsep}{0pt}\begin{tabular}{c}\Large{\#\ First\ Stage\ Triggers\ (\ $\times10^6$\ )}\end{tabular}}%
\psfrag{s07}[t][t]{\color[rgb]{0,0,0}\setlength{\tabcolsep}{0pt}\begin{tabular}{c}\Large{Total\ Mass\ /$M_\odot$}\end{tabular}}%
\psfrag{s08}[l][l]{\color[rgb]{0,0,0}Tapered}%
\psfrag{s45}[l][l]{\color[rgb]{0,0,0}\Large{Untapered}}%
\psfrag{s46}[l][l]{\color[rgb]{0,0,0}\Large{Tapered}}%
%
\psfrag{x01}[t][t]{0}%
\psfrag{x02}[t][t]{0.1}%
\psfrag{x03}[t][t]{0.2}%
\psfrag{x04}[t][t]{0.3}%
\psfrag{x05}[t][t]{0.4}%
\psfrag{x06}[t][t]{0.5}%
\psfrag{x07}[t][t]{0.6}%
\psfrag{x08}[t][t]{0.7}%
\psfrag{x09}[t][t]{0.8}%
\psfrag{x10}[t][t]{0.9}%
\psfrag{x11}[t][t]{1}%
\psfrag{x12}[t][t]{\Large{20}}%
\psfrag{x13}[t][t]{\Large{30}}%
\psfrag{x14}[t][t]{\Large{40}}%
\psfrag{x15}[t][t]{\Large{50}}%
\psfrag{x16}[t][t]{\Large{60}}%
\psfrag{x17}[t][t]{\Large{70}}%
\psfrag{x18}[t][t]{\Large{80}}%
\psfrag{x19}[t][t]{\Large{90}}%
\psfrag{x20}[t][t]{\Large{100}}%
%
\psfrag{v01}[r][r]{0}%
\psfrag{v02}[r][r]{0.1}%
\psfrag{v03}[r][r]{0.2}%
\psfrag{v04}[r][r]{0.3}%
\psfrag{v05}[r][r]{0.4}%
\psfrag{v06}[r][r]{0.5}%
\psfrag{v07}[r][r]{0.6}%
\psfrag{v08}[r][r]{0.7}%
\psfrag{v09}[r][r]{0.8}%
\psfrag{v10}[r][r]{0.9}%
\psfrag{v11}[r][r]{1}%
\psfrag{v12}[r][r]{\Large{1}}%
\psfrag{v13}[r][r]{\Large{2}}%
\psfrag{v14}[r][r]{\Large{3}}%
\psfrag{v15}[r][r]{\Large{4}}%
\psfrag{v16}[r][r]{\Large{5}}%
\psfrag{v17}[r][r]{\Large{6}}%
\psfrag{v18}[r][r]{\Large{7}}%
%
\resizebox{6cm}{!}{\includegraphics{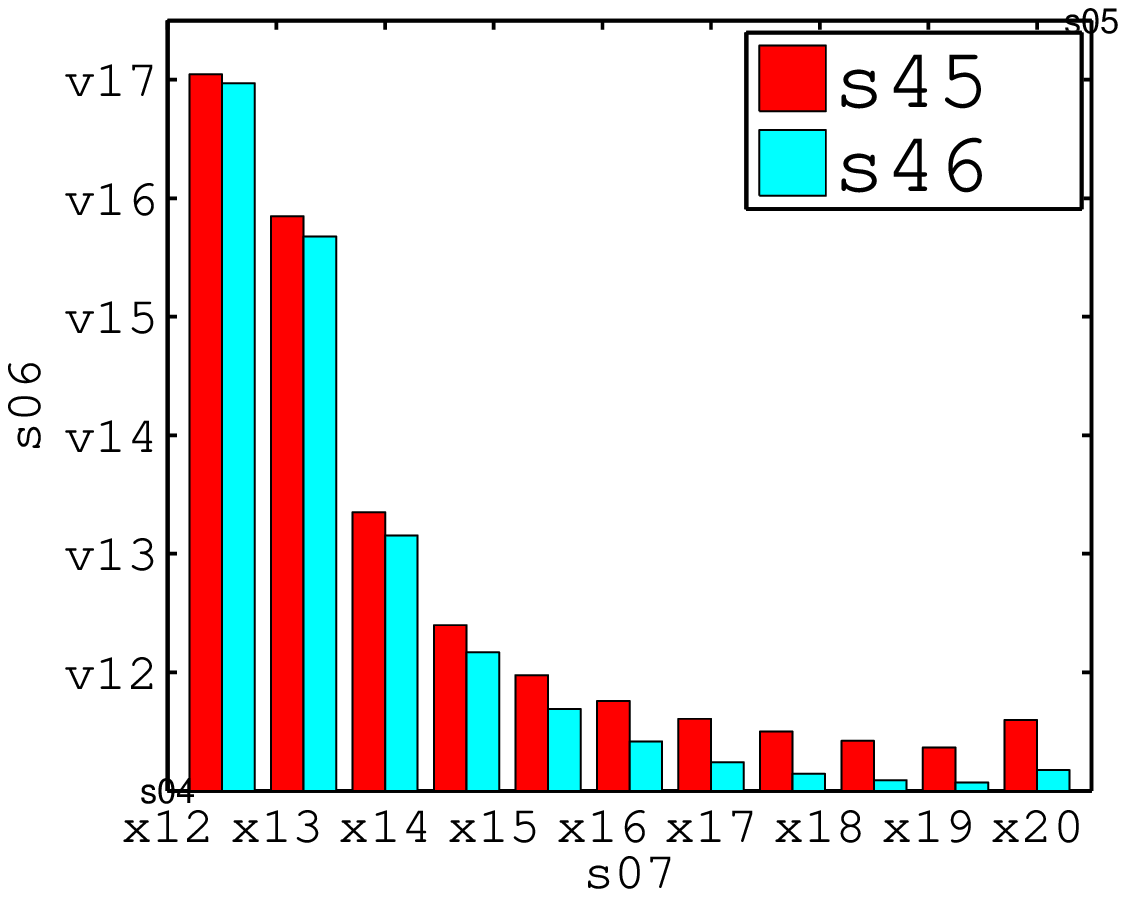}}%
\end{psfrags}%
%

\end{minipage}
\hspace{2.25cm}
\begin{minipage}[h]{4cm}
\vspace{0.2cm}
%
%
\begin{psfrags}%
\psfragscanon%
%
\psfrag{s01}[b][b]{\color[rgb]{0,0,0}\setlength{\tabcolsep}{0pt}\begin{tabular}{c}\Large{\#\ Second\ Stage\ Coincident\ Triggers\ (\ $\times10^3$\ )}\end{tabular}}%
\psfrag{s05}[][]{\color[rgb]{0,0,0}\setlength{\tabcolsep}{0pt}\begin{tabular}{c} \end{tabular}}%
\psfrag{s06}[][]{\color[rgb]{0,0,0}\setlength{\tabcolsep}{0pt}\begin{tabular}{c} \end{tabular}}%
\psfrag{s07}[t][t]{\color[rgb]{0,0,0}\setlength{\tabcolsep}{0pt}\begin{tabular}{c}\Large{Total\ Mass\ /$M_\odot$}\end{tabular}}%
\psfrag{s08}[l][l]{\color[rgb]{0,0,0}Tapered}%
\psfrag{s33}[l][l]{\color[rgb]{0,0,0}\Large{Untapered}}%
\psfrag{s34}[l][l]{\color[rgb]{0,0,0}\Large{Tapered}}%
%
\psfrag{x01}[t][t]{0}%
\psfrag{x02}[t][t]{0.1}%
\psfrag{x03}[t][t]{0.2}%
\psfrag{x04}[t][t]{0.3}%
\psfrag{x05}[t][t]{0.4}%
\psfrag{x06}[t][t]{0.5}%
\psfrag{x07}[t][t]{0.6}%
\psfrag{x08}[t][t]{0.7}%
\psfrag{x09}[t][t]{0.8}%
\psfrag{x10}[t][t]{0.9}%
\psfrag{x11}[t][t]{1}%
\psfrag{x12}[t][t]{\Large{20}}%
\psfrag{x13}[t][t]{\Large{30}}%
\psfrag{x14}[t][t]{\Large{40}}%
\psfrag{x15}[t][t]{\Large{50}}%
\psfrag{x16}[t][t]{\Large{60}}%
\psfrag{x17}[t][t]{\Large{70}}%
\psfrag{x18}[t][t]{\Large{80}}%
\psfrag{x19}[t][t]{\Large{90}}%
\psfrag{x20}[t][t]{\Large{100}}%
%
\psfrag{v01}[r][r]{0}%
\psfrag{v02}[r][r]{0.1}%
\psfrag{v03}[r][r]{0.2}%
\psfrag{v04}[r][r]{0.3}%
\psfrag{v05}[r][r]{0.4}%
\psfrag{v06}[r][r]{0.5}%
\psfrag{v07}[r][r]{0.6}%
\psfrag{v08}[r][r]{0.7}%
\psfrag{v09}[r][r]{0.8}%
\psfrag{v10}[r][r]{0.9}%
\psfrag{v11}[r][r]{1}%
\psfrag{v12}[r][r]{\Large{2}}%
\psfrag{v13}[r][r]{\Large{4}}%
\psfrag{v14}[r][r]{\Large{6}}%
\psfrag{v15}[r][r]{\Large{8}}%
\psfrag{v16}[r][r]{\Large{10}}%
%
\resizebox{6cm}{!}{\includegraphics{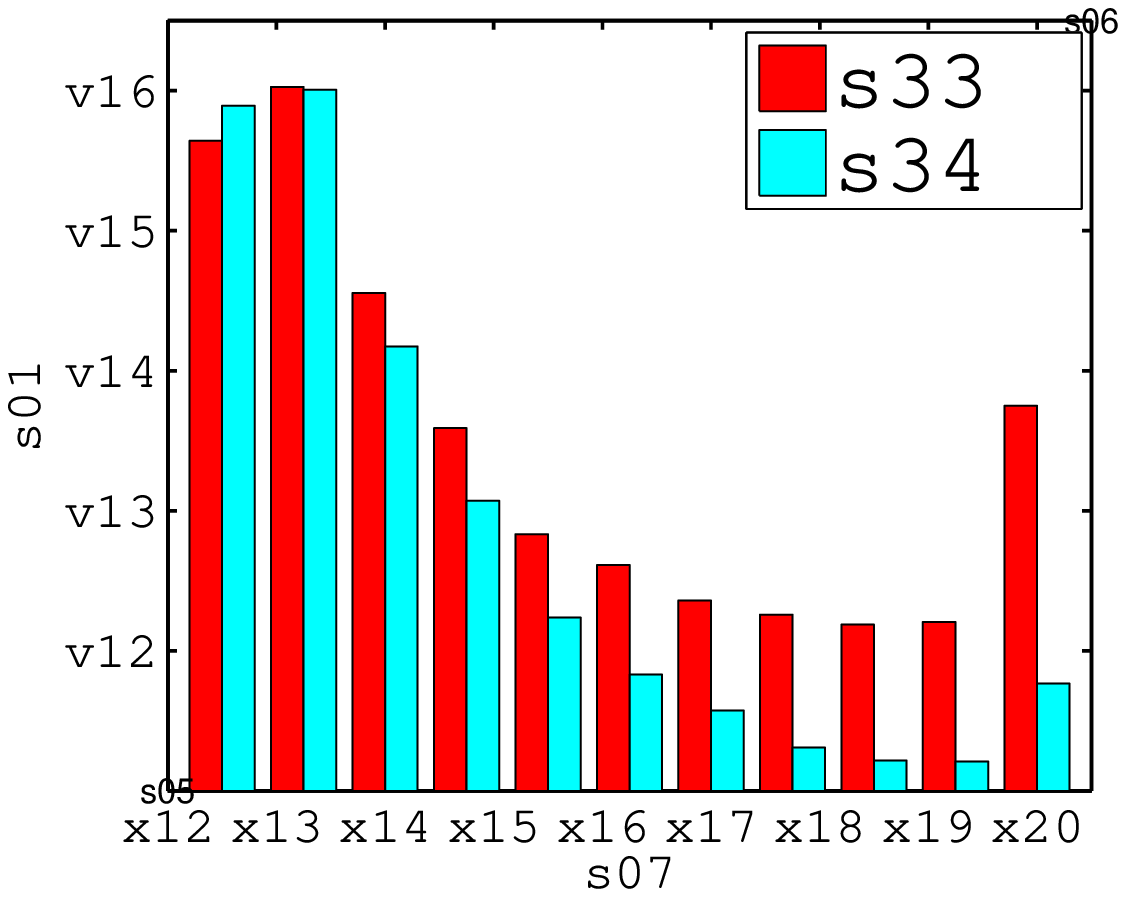}}%
\end{psfrags}%
%

\end{minipage}
\caption{\label{fig:triggers}
\textit{Left}: Number of triggers recovered by matched filtering the S4 data 
with and without tapering applied to the templates. 
\textit{Right}: Number of triggers at the second stage of the analysis 
pipeline, after consistency checks and coincidence tests \cite{Robinson:2008}
in the time-of-arrival and masses of the component stars have been applied.
}
\end{figure}

Figure~\ref{fig:triggers} shows the number of triggers as a function of
total mass with and without tapering. It can be seen that the number of 
triggers is generally higher when the templates are not tapered. The only 
exception seems to be the lowest mass bin in the second-stage coincident 
triggers, where the opposite is true. 
However, the difference in the number of triggers in this bin is not large, 
and is likely just a statistical anomaly. 
For first-stage single-detector
triggers, the number of triggers using tapered templates is $84\%$
of that obtained using un-tapered templates. The number of second-stage 
coincident triggers when using tapered templates is $71\%$ of 
that obtained for un-tapered templates.
The difference in trigger rates is more significant
at higher masses. This is because the template waveforms for these systems 
terminate at a frequency within or below the most sensitive frequency band 
of the detector, making any leakage of power to higher frequencies more 
significant (cf Figure \ref{fig:spectra}, left most panel). The reduced 
trigger rate indicates that applying the taper 
function to the templates could aid in reducing the false alarm rate in a 
search for high mass \ac{CBC}s.

\section{Effect of windowing on detection efficiency and parameter estimation}
\label{sec:pe}
The same data used in section \ref{sec:triggers} were re-analysed, but with
simulated gravitational wave signals (injections) added.
The injections were of the same family as the templates used in section
~\ref{sec:triggers}. This allowed us to compare the detection 
efficiencies and accuracy of parameter estimation using tapered vs. untapered 
templates. 
We looked at the error in recovered chirp mass and arrival time at both
single detector first stage triggers and coincident second stage triggers, but
found negligible difference between the two cases. \footnote{ We have seen 
some evidence of improvements in parameter estimation for the ambiguity 
function of high mass inspiral-only waveforms, but this is outside the
context of a gravitational wave search.} 

We did not explicitly measure the detection efficiency as a function of
distance, but found the number of injections recovered to be nearly identical
in the two cases, with less than $1\%$ fewer injections found when using tapered 
templates.  Given the vast reduction in the trigger rates shown in 
Section~\ref{sec:triggers}, this indicates that an improvement in detection 
efficiency \textit{can be expected} when using tapered templates.

The above studies were performed first with tapering applied to the 
injections and then repeated without - the difference between the
results was negligible.

\section{Conclusion}\label{sec:conc}

We have developed a tapering method that leads to a spectrum for \ac{TD}
waveforms that more closely matches their \ac{FD} analogs, containing
significantly less power at unexpected frequencies when compared with the use
of a square window. This is acheived by automating the implementation of the
window.

If tapering is applied to templates in a 
gravitational wave search the trigger rates are reduced, especially for
high mass templates, without any significant change in detection efficiency.
In a search, foreground triggers can be ranked by their probability of
occurring as a background trigger; thus if background triggers are reduced, 
a given foreground trigger may appear more significant.
Another benefit of reduced trigger rates is that the computational cost of a
search will decrease. We have demonstrated that the windowing method would be
beneficial when used in a high mass search.

The tapering method could also be useful in low latency data analysis
techniques where \ac{TD} templates
are divided into sub-templates of different frequency ranges, and 
matched filtered individually~\cite{Acernese:2006uu}. The relative shortness of
some templates in the higher frequency bands potentially compounds
the problem of using a square window, and tapering the templates may go
some way to alleviating this issue.

\ack
We thank the \ac{LSC} for their permission to use the \ac{S4} data, 
without which this investigation would not have been possible. We would 
also like to thank Evan Ochsner and Anand Sengupta for their efforts in 
some of the inital tapering studies and the members of the \ac{CBC} search 
group of the \ac{LSC}-Virgo collaboration for allowing us to use the 
\ac{CBC} search pipeline. We are grateful to Jolien Creighton for
his critical analysis, which allowed us to clarify and improve this paper.
We also thank Thomas Dent for carefully reading
the manuscript and for his helpful suggestions. This work was funded in part 
by STFC studentship PPA/S/S/2006/4330 and STFC grant PP/F001096/1.

\section*{References}
\bibliographystyle{iopart-num}
\bibliography{iulpapers}

\end{document}